# STRIPLINE KICKER FOR INTEGRABLE OPTICS TEST ACCELERATOR

Sergey A. Antipov, University of Chicago, Chicago, Illinois,
Alexander Didenko, Valeri Lebedev, Alexander Valishev, Fermilab, Batavia, Illinois

*Abstract*

We present a design of a stripline kicker for Integrable Optics Test Accelerator (IOTA). For its experimental program IOTA needs two full-aperture kickers, capable to create an arbitrary controllable kick in 2D. For that reason their strengths are variable in a wide range of amplitudes up to 16 mrad, and the pulse length 100 ns is less than a revolution period for electrons. In addition, the kicker should have a physical aperture of 40 mm for a proposed operation with proton beam, and an outer size of 70 mm to fit inside existing quadrupole magnets to save space in the ring. Computer simulations using CST Microwave Studio® show high field uniformity and wave impedance close to 50 Ω.

## KICKER DESCRIPTION

Integrable Optics Test Accelerator is a relatively small research storage ring with the circumference of 40 m. It will operate with short bunches of 150 MeV electrons injected from the ASTA linac [1], and later with 2.5 MeV protons from HINS RFQ [2]. The ring has a flexible linear optics to allow for a variety of physics experiments [3], including nonlinear integrable optics, the concept of which is described in [4]

The nonlinear dynamics research program at IOTA involves construction of Poincare maps of phase space, and that requires two kickers: a horizontal and a vertical to be able to place the beam at an arbitrary initial amplitudes in the transverse phase space. The kickers have to be tunable in a wide range from small to full-aperture, with the physical aperture of the machine being quite large: beam pipe inner radius is 24 mm or 40 sigma of horizontal size of electron beam at the place with maximum beta-function [3]. In addition, the vertical kicker will also serve as an injection kicker.

To support the required short length of kicker pulse a stripline kicker was preferred. It is also relatively simple in design and fabrication. The kickers are going to be fed by recommissioned power supplies from Fermilab's Tevatron injection kickers, which are capable of producing variable voltage pulses up to 25 kV. The power supplies have rise/fall time of about 20 ns, and are capable of producing a 100 ns-long pulse, which is slightly shorter than a revolution period of the ring – 130 ns. That allows us to make sure that the beam is affected on one turn only, and that is critical for the accuracy of experiments. The kickers are loaded with 50 Ω external resistive load.

Position of kickers in the ring is depicted in Fig. 1. Both kickers are located in the injection straight, downstream to injection Lambertson magnet. To minimize the required space one of the kickers is placed inside a quadrupole doublet. Their aperture of 72 mm determines a limit on the outer dimensions of the kicker's pipe.

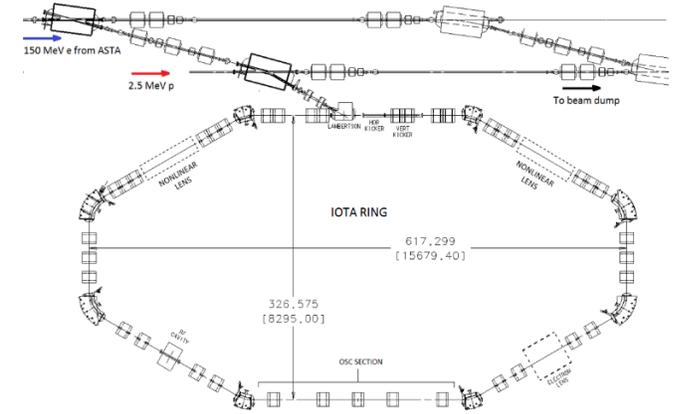

Figure 1: IOTA Ring, configuration for nonlinear optics.

Table 1: Parameters of IOTA ring and its kickers for electron operation. Proton parameters are in parentheses.

| Beam energy | 150 MeV (2.5 MeV) |
|---|---|
| RMS beam size at kicker | ~ 0.2 mm  (~ 3 mm) |
| Max plate voltage | ±25 kV |
| Pulse duration, rep. rate | 100 ns, < 1 Hz |
| Max. kick angle: hor., vert. | 16, 8 mrad |

## DESIGN

The design of the kickers was inspired by a work carried out at the Budker Institute (BINP), Russia [5]. The geometry was adjusted to fit inside the existing quadrupole magnets. Since IOTA ring hosts several different experiments, requiring their own set of optics, it was crucial to minimize limitation of physical aperture, while meeting the requirements on the strength, field quality, wave impedance, and spatial constraints. The physical aperture was tentatively set to be at least 40 mm for future experiments with a proton beam. The kicker cross-section is depicted in Fig. 2.

The kicker has two "banana"-shaped plates, separated by 40 mm. Each plate is extruded from Al and is 6 mm-thick. Such thickness is required to prevent bending of the plates under their own weight and twisting during the fabrication and installation. The pipe is elliptical, with one of the half axes only slighter (4 mm) longer that the other. This shape provides lower electric field near the edges, than the circular shape of the same wave impedance, and it complies with the spatial constraints.

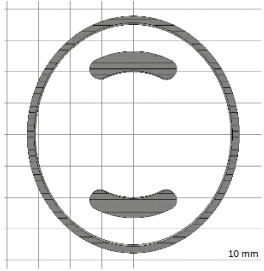

Figure 2: Kicker cross-section (vertical kicker).

Although the chamber is not exactly circular, wave impedance of the kicker is approximately described by the formula:

$$Z = \frac{Z_0}{\theta} \ln \frac{r_{pipe}}{r_{line}}, \qquad (1)$$

where $Z_0$ is vacuum impedance and $\theta$ – opening angle; and the strength of the kicker is:

$$\alpha_{max} = \frac{2V_{max} L_{kick}}{r_{line}} g(\omega), \qquad (2)$$

where $g(\omega) = \frac{4}{\pi} \sin \frac{\theta}{2}$ is the geometric factor.

The plates are supported by ceramic inserts, placed roughly every 50 cm. The ceramic inserts are attached to the plates in such way that there is no direct line of sight from the beam (the beam does not "see" the ceramics). This is done to prevent possible charge accumulation on the surface of the inserts, introduced by secondary electrons, created by the beam in residual gas. Figure 3 shows a model of the vertical kicker inside a quadrupole doublet; and Table 2 summarizes the main geometry parameters of the kickers. The assembly does not have its own mechanical support, and is supported on the flanges.

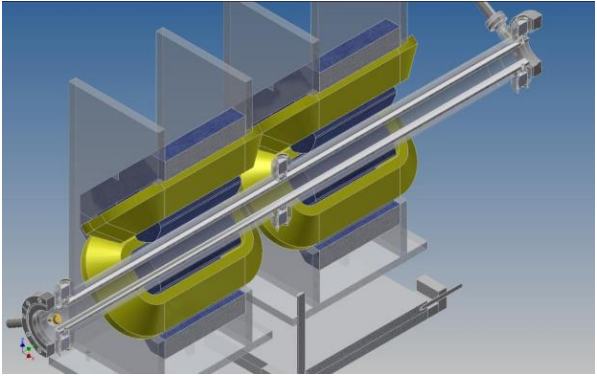

Figure 3: Vertical kicker inside a quadrupole doublet.

## SIMULATION RESULTS

E&M field simulations were performed in CST Studio® suite to check field magnitude and wave impedance of the kickers. Figure 4 shows a 3D picture of E-field distribution in the vertical kicker, when a voltage of ±25 kV is applied to the plates. The E-field does not exceed 50 kV/cm near the edges and 15 kV/cm along ceramics. A distribution of electric field in a transverse cross-section is shown in Fig. 5, and Fig. 6 presents a corresponding plot of electric field dependence along vertical and horizontal axes.

Table 2: Basic geometry of vertical (horizontal) kicker

| | |
|---|---|
| Vacuum pipe half axes, H/V | 31/ 35 mm |
| Vacuum pipe thickness, material | 2 mm, stainless steel |
| Plate inner radius | 20 mm |
| Plate thickness, material | 6 mm, Al |
| Plate opening angle | 65 deg. |
| Length of plates | 1050 (580) mm |
| Total length | 1105 (635) mm |
| Number of ceramic inserts | 3 (2) |

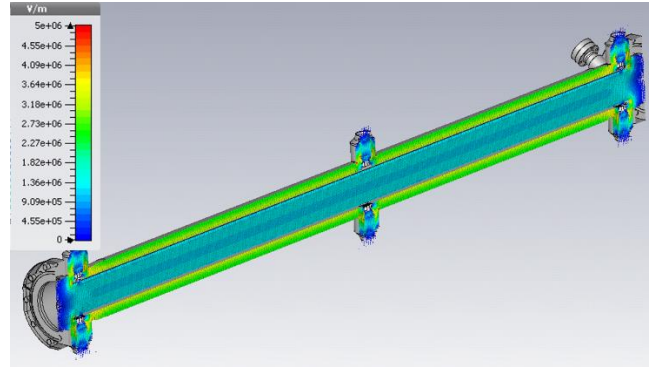

Figure 4: Electric field distribution (vert. kicker).

Signal voltage reflections due to impedance mismatch can be calculated as $\Gamma = (Z - Z_0)/(Z + Z_0)$, where $Z_0 = 50\Omega$ is the matched impedance, and shall not exceed 5% for both modes of operation: with voltage supplied to both plates or only one.

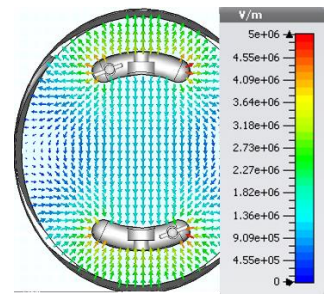

Figure 5: Transverse cross-section of electric field (vert. kicker).

According to Figs. 5, 6, the field variations do not exceed 5% on the radius equal to the RMS size of the injected proton beam of 3 mm. Presently it is considered to be sufficient for the operation with proton beam. The size of electron beam is significantly smaller (0.2 mm) and field non-uniformity does not represent a problem for operations with electrons.

If required a further improvement of field quality is possible for operation with protons. Low proton velocity,

$\beta = 0.073$, requires about 7 times lower voltage than for electrons. It makes possible a replacement of the plates with wider and flatter ones resulting in a better field uniformity.

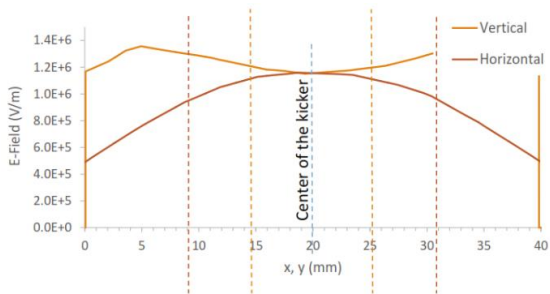

Figure 6: Kick strength as a function of displacement from the central axis along a horizontal and vertical lines. Dotted lines represent 3 sigma size of the proton beam: orange – vertical, red - horizontal

Table 3: Simulation results

| | |
|---|---|
| Max voltage | ± 25 kV |
| Max E-field: volume, along ceramics | 50 kV/cm, 15 kV/cm |
| E-field on the axis | 11.5 kV/cm |
| Wave impedance: dipole, sum mode | 45, 50 Ω |

## CONCLUSION

A unified design has been created for IOTA's horizontal and vertical kickers. The kickers will serve for beam injection to the ring and its experimental program. Their design satisfies the requirements on strength, field quality, wave impedance, and physical aperture. In addition, it allows to save space in the machine by placing a kicker inside a quadrupole. The kickers provide a low-cost solution, allowing to reuse existing power supplies from Tevatron. The same kickers can be used for operation with a proton beam on the second stage of the experimental program, although this will require another pulse generator.

## ACKNOWLEDGEMENT

Fermilab is operated by Fermi Research Alliance, LLC under Contract No. DE-AC02-07CH11359 with the United States Department of Energy.